\documentclass[11pt]{article}

\usepackage{mymacros}

\begin{document}

\floatpagestyle{plain}

\pagenumbering{roman}

\renewcommand{\headrulewidth}{0pt}
\rhead{
OHSTPY-HEP-T-15-003
}
\fancyfoot{}

\title{\huge \bf{A Pati-Salam Version of \\Subcritical Hybrid Inflation}}

\author{B. Charles Bryant and Stuart Raby}

\affil{\em Department of Physics, The Ohio State University,\newline
191 W.~Woodruff Ave, Columbus, OH 43210, USA \enspace\enspace\enspace\enspace \medskip}

\date{}

\maketitle
\thispagestyle{fancy}

\begin{abstract}\normalsize\parindent 0pt\parskip 5pt
In this paper we present a model of subcritical hybrid inflation with a Pati-Salam [PS] symmetry group. Both the inflaton and waterfall fields contribute to the necessary e-foldings of inflation, while only the waterfall field spontaneously breaks PS hence monopoles produced during inflation are diluted during the inflationary epoch. The model is able to produce a tensor-to-scalar ratio, $r < 0.09$ consistent with the latest BICEP2/\emph{Keck} and Planck data, as well as scalar density perturbations and spectral index, $n_s$, consistent with Planck data.  For particular values of the parameters, we find $r = 0.084$ and $n_s = 0.0963$. The energy density during inflation is directly related to the PS breaking scale, $v_{PS}$.  The model also incorporates a $\mathbb{Z}_4^R$ symmetry which can resolve the $\mu$ problem and suppress dimension 5 operators for proton decay, leaving over an exact $R$-parity. Finally the model allows for a complete three family extension with a $D_4$ family symmetry which reproduces low energy precision electroweak and LHC data.
\end{abstract}

\clearpage
\newpage

\pagenumbering{arabic}

\section{Introduction}

The Pati-Salam (PS) gauge symmetry,  $SU(4)_C \times SU(2)_L \times SU(2)_R$, has the nice feature that it unifies one family of quarks and leptons into two irreducible representations, $ Q = (4, 2, 1) \supset \{q, \; \ell \}, \;  Q^c = (\bar 4, 1, \bar 2) \supset \{ (\begin{array}{c}  u^c \\ d^c \end{array}), \; (\begin{array}{c} \nu^c \\ e^c \end{array}) \}.$   In addition, the two Higgs doublets of the minimal supersymmetric standard model (MSSM) appear in one irreducible representation of PS given by ${\cal H} = (1, 2, \bar 2)$.  This allows for the possibility of Yukawa unification for the third generation of quarks and leptons with one universal coupling given by $\lambda Q_3 \ {\cal H} \ Q^c_3$ with $\lambda_t = \lambda_b = \lambda_\tau = \lambda_{\nu_\tau} \equiv \lambda$ at the GUT scale.  Although PS does not unify all the gauge couplings, it is possible that the PS gauge symmetry is the four dimensional gauge symmetry resulting from a 5D or 6D orbifold GUT such as SO(10).   In this case, gauge coupling unification is enforced by the higher dimensional unification.   In fact, it has been shown that PS gauge symmetry in 4D can be obtained from heterotic orbifold constructions~\cite{Kobayashi:2004ud,Kobayashi:2004ya}.

In this paper we discuss inflationary dynamics governed by subcritical hybrid inflation~\cite{Buchmuller:2014rfa,Buchmuller:2014dda,Wieck:2014xxa,Kohri:2013gva} with a waterfall field which spontaneously breaks the PS symmetry.\footnote{In contrast to Refs. \cite{Buchmuller:2014rfa,Buchmuller:2014dda} we use $F$-term inflation instead of $D$-term inflation.} In the subcritical hybrid inflation scenario, the coupling between the inflaton and the waterfall field is sufficiently small so that the stage of inflation after the critical point may last for more than 60 $e$-folds. The value of the inflaton at the critical point can therefore be large relative to the Planck mass allowing for GUT-scale inflation. After the critical point, the waterfall field quickly settles into an inflaton-dependent minimum which in turn yields an effective single-field inflaton potential. The potential is essentially an interpolation between a nearly flat potential at large field values and a quadratic potential at low field values. Such an arrangement allows for a tensor-to-scalar ratio prediction between that of traditional hybrid inflation and chaotic inflation. Furthermore, we are able to directly identify the energy scale during inflation with the PS/GUT breaking scale and at the same time obtain a tensor-to-scalar ratio, $r \sim 0.08$.  This was also accomplished in Refs.~\cite{Buchmuller:2014rfa,Buchmuller:2014dda,Wieck:2014xxa} where the energy scale during inflation was associated with a $B - L$ breaking scale.\footnote{In previous models in the literature, Refs.~\cite{Dvali:1994ms,Jeannerot:2000sv,Antusch:2010va,Buchmuller:2013dja}, the energy scale during inflation was also related to a GUT symmetry breaking scale.  However, due to the fact that the inflaton potential during inflation was very flat, the tensor-to-scalar ratio $r < 10^{-2}$.}

Since the waterfall occurs before the last 60 $e$-foldings of inflation, the monopoles produced by the PS symmetry breaking \cite{Lazarides:1980cc} are diluted away during inflation. Moreover they are not produced after reheating. Such a solution to the monopole problem has been presented previously in the context of hybrid inflation \cite{Jeannerot:2000sv,Antusch:2010va}, although the details here are markedly different. The model also has a \zfr symmetry which
can be dynamically broken to solve the $\mu$ problem and eliminate problems with dimension 5 operators mediating proton decay. The resulting low energy theory retains an exact $R$-parity. In addition we show how to obtain a 3 family model for quark and lepton masses, with a $D_4$ family symmetry, which is known to be consistent with low energy data. Finally we discuss reheating in the model, however we defer to work in progress for discussions of leptogenesis, dark matter, and a potential gravitino problem.


\section{Inflation Sector}

\subsection{Model}

The superpotential and K\"{a}hler potential for the inflaton sector of the model with an $SU(4)_C \times SU(2)_L \times SU(2)_R$ gauge symmetry times \zfr discrete $R$ symmetry are given by
\bea
\fw_{I} &=&
 \Phi \ \left( \k \ \Sb^c S^c + m \ Y + \alpha \ {\cal H}  {\cal H}\right) +  \l X\lrp{\Sb^c S^c-\frac{v^2_{PS}}{2}}
     + S^c \Sigma S^c + \Sb^c \Sigma \Sb^c \\
\fk &=& \half(\Phi+\Phi^\dag)^2 + (S^c)^\dag S^c + (\Sb^c)^\dag \Sb^c + Y^\dag Y+
 X^\dag X\lrs{1-c_X \frac{X^\dag X}{\mpl^2}+a_X\lrp{\frac{X^\dag X}{\mpl^2}}^2 }
,\eea
where the inflaton and waterfall superfields are given by $ \{\Phi = (1, 1, 1, 2), \; S^c = (\bar{4}, 1, 2, 0), \; \Sb^c = (4, 1, 2, 0) \} $. As a consequence the Pati-Salam gauge symmetry is broken to the Standard Model at the waterfall transition and remains this way both during inflation and afterwards. The superfield, $\Sigma = (6,1,1,2)$, is needed to guarantee that the effective low energy theory below the PS breaking scale is just the MSSM. The singlet $X = (1,1,1,2)$ is introduced in order to obtain $F$-term hybrid inflation in which the coupling of the inflaton to the waterfall field is independent of the self-coupling of the waterfall field. The term with the singlet $Y=(1,1,1,0)$ is added in order to obtain a supersymmetric vacuum after inflation. The parameter $m$ is smaller than in typical chaotic inflation models and the $F$-term of $Y$ acts to lift the flatness of the potential above the critical point.  The term with the Higgs field, ${\cal H}= (1, 2, \bar 2, 0)$, is added to enable reheating.  This will be discussed later. The K\"{a}hler potential has a shift symmetry, $\im{\Phi} \ra \im{\Phi} + \Theta$, where $\Theta$ is a real constant. The constant $c_X$ is necessary for $X$ to have a mass larger than the Hubble parameter during inflation, so that during inflation the field is stabilized at zero, and the constant $a_X$ is necessary for the potential to be bounded from below~\cite{Carpenter:2014saa}.

\subsection{Inflationary dynamics}

We consider now the inflationary dynamics in this theory. Due to the shift symmetry in the K\"{a}hler potential,  $\im{\Phi}$ can take trans-Planckian values without causing the scalar potential to become very steep, and so we identify this field as the inflaton. We assume that at the waterfall transition the fields $S^c, \Sb^c$  obtain vevs in the $\nu^c$ direction. Following Buchm\"{u}ller et al.~\cite{Buchmuller:2012wn}, we express the waterfall fields in the unitary gauge so that the physical degrees of freedom are manifest in the subsequent treatment of reheating. The superfields $S^c, \ \Sb^c$ are written as
\bea
S^c &=&  \exp(i \ T_{SU(4)_C/SU(3)} \ \phi_u^c) \ \exp(i \ T_{SU(2)_R/U(1)_R} \ \phi_e^c) \
\lrp{\begin{array}{cc} 0 & \frac{1}{\sqrt{2}}\exp(i \ T) \ V^c \\ d_S^c & 0 \end{array}}
\nn\\
\Sb^c &=&  \exp(-i \ T^T_{SU(4)_C/SU(3)} \ \bar \phi_u^c) \ \exp(-i \ T^T_{SU(2)_R/U(1)_R} \  \bar \phi_e^c) \
\lrp{\begin{array}{cc} 0 & \frac{1}{\sqrt{2}}\exp(-i \ T) \ V^c \\ \bar d_{\Sb}^c & 0 \end{array}}
.\eea

The fields  $\phi_u^c, \ \bar \phi_u^c, \ \phi_e^c, \ \bar \phi_e^c, \ T$ are goldstone fields which are eaten by the broken $SU(4)_C$ and $SU(2)_R$ supergauge fields.  The gauge bosons in $(SU(4)_C \times SU(2)_L \times SU(2)_R)/(SU(3) \times SU(2) \times U(1))$ obtain masses of order $\frac{g}{2} v_{PS}$, where $g = g_4 \approx g_{2R}$ is of order 1. The fields $d_S^c, \ \bar d_{\Sb}^c$ get a supersymmetric mass with the color triplets in $\Sigma$ and the scalar component of $V^c$ ($s$) gets a vev breaking PS to the SM. Thus Pati-Salam is spontaneously broken at the waterfall transition and remains broken after inflation. As a result any monopole density formed during the breaking of Pati-Salam to the SM \cite{Lazarides:1980cc} is diluted during inflation.

The $F$-terms in the global SUSY limit are
\be
F_\Phi=\frac{\k}{2} (V^c)^2+mY \;,\quad
F_{V^c} = \lrp{\k\Phi+\l X }V^c \;,\quad
F_X = \frac{\l}{2}\lrp{(V^c)^2-v^2_{PS}}  \;,\quad
F_Y = m\Phi
\;.\ee
Before inflation, only $\Phi$ has a nonzero vev and only $F_\Phi$ and $F_{V^c}$ vanish. Thus supersymmetry is broken before and during inflation. After inflation, $\vev{\Phi}$ goes to zero and $\vev{V^c}=v_{PS}$ and $\vev{Y}=-\k v^2_{PS}/(2m)$, restoring supersymmetry. $X$ remains stabilized at zero throughout.
The $D$-term scalar potential is given by
\be
V_D = \sum_a \frac{g_a^2}{2} \ \lrp{(S^c)^* \ T_a \ S^c - \Sb^c \ T_a \ (\Sb^c)^* + \cdots}^2
,\ee
where $T_a$ are the generators of PS in the $(\bar 4, 1, \bar 2)$ representation and $V_D = 0$ during inflation.

The real scalar components of the inflaton, waterfall, and $Y$ superfields may be expressed as
\be
\Phi\supset\frac{a+i\phi}{\sqrt{2}} \;,\quad
V^c \supset \frac{s+i\tau}{\sqrt{2}} \;,\quad
Y \supset \frac{y + i u}{\sqrt{2}}
\;.
\ee

Before the waterfall transition, the fields $a$, $s$, $\tau$, $y$, and $u$ have positive squared masses and are stabilized at the origin. Once the inflaton reaches subcritical field values, the field $s$ develops a tachyonic mass and the waterfall transition is triggered. We represent the symmetry breaking in the Lagrangian by the field shifts, $s=\s+\sqrt{2}v_{PS}$ and $y = h - \k v_{PS}^2/\sqrt{2}m$. In our setup, the coupling between the inflaton and the waterfall field, $\k$, is taken to be much smaller than the waterfall field self-coupling, $\l$. Furthermore, the parameter $m$ will be taken to be $10^{-6}\mpl$ and the PS scale, $v_{PS}$ will be $\sim10^{-2}\mpl$. It has been shown that in this scenario the proceeding stage of tachyonic preheating occurs for a few $e$-folds but produces kinetic and gradient energy that is severely subdominant to the vacuum energy and therefore fails to terminate inflation~\cite{Buchmuller:2014rfa}. In fact a considerably large number of $e$-folds can still be generated after symmetry breaking.

At subcritical field values, the inflationary dynamics are determined by the $F$-term supergravity scalar potential,
\begin{align}
V(\phi,s,y)
&\simeq  \frac{\lam^2 v_{PS}^4 }{4} +
    \frac{1}{4}\lrs{\lrp{\frac{m^2}{\mpl^2}+\k^2} \phi^2 - \lam^2 v_{PS}^2 }s^2 + 
    \frac{\lrp{\lam^2+\k^2}}{16}s^4 +
    \half m^2\f^2 + \frac{\k m}{2\sqrt{2}} s^2  y
\nn\\
&\quad\;\;   
    +\frac{1}{4}\lrs{2\lrp{m^2 + \frac{\l^2 v^4_{PS}}{4\mpl^2}}+\lrp{m^2-\frac{\l^2v^2_{PS}}{2}}\frac{s^2}{\mpl^2} + \frac{\l^2}{8} \frac{s^4}{\mpl^2}}y^2 +
    \frac{1}{4}\lrp{\frac{m^2}{\mpl^2}+\frac{\l^2v^4_{PS}}{8\mpl^4}}y^4 
.
\label{eq:vinf}
\end{align}

\begin{figure}[t]
\centering
\begin{subfigmatrix}{2}
\subfigure{
\includegraphics[width=0.45\textwidth, trim=0ex 0ex 0ex 0ex, clip]{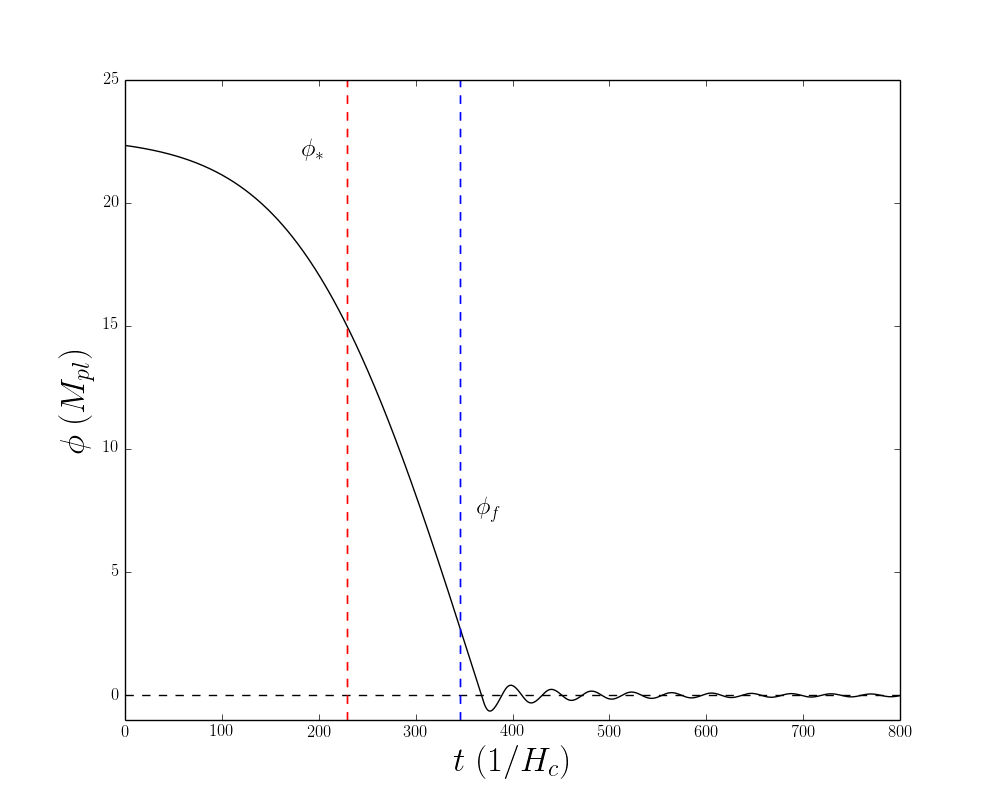}}
\subfigure{
\includegraphics[width=0.45\textwidth, trim=0ex 0ex 0ex 0ex, clip]{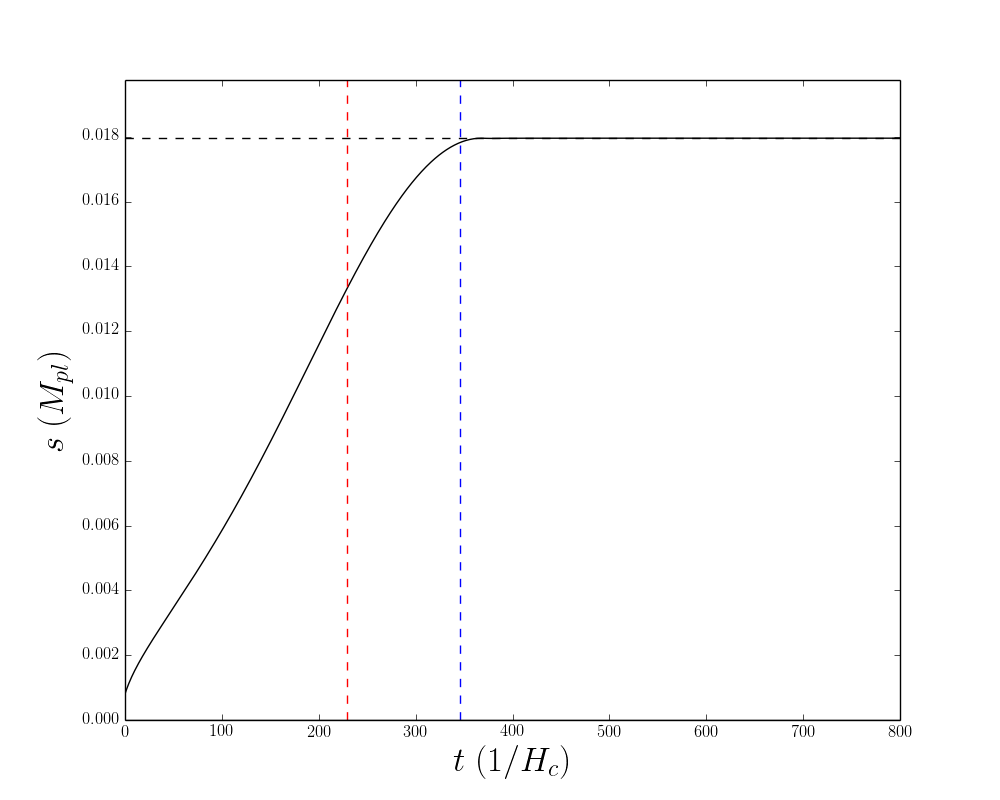}}
\subfigure{
\includegraphics[width=0.45\textwidth, trim=0ex 0ex 0ex 0ex, clip]{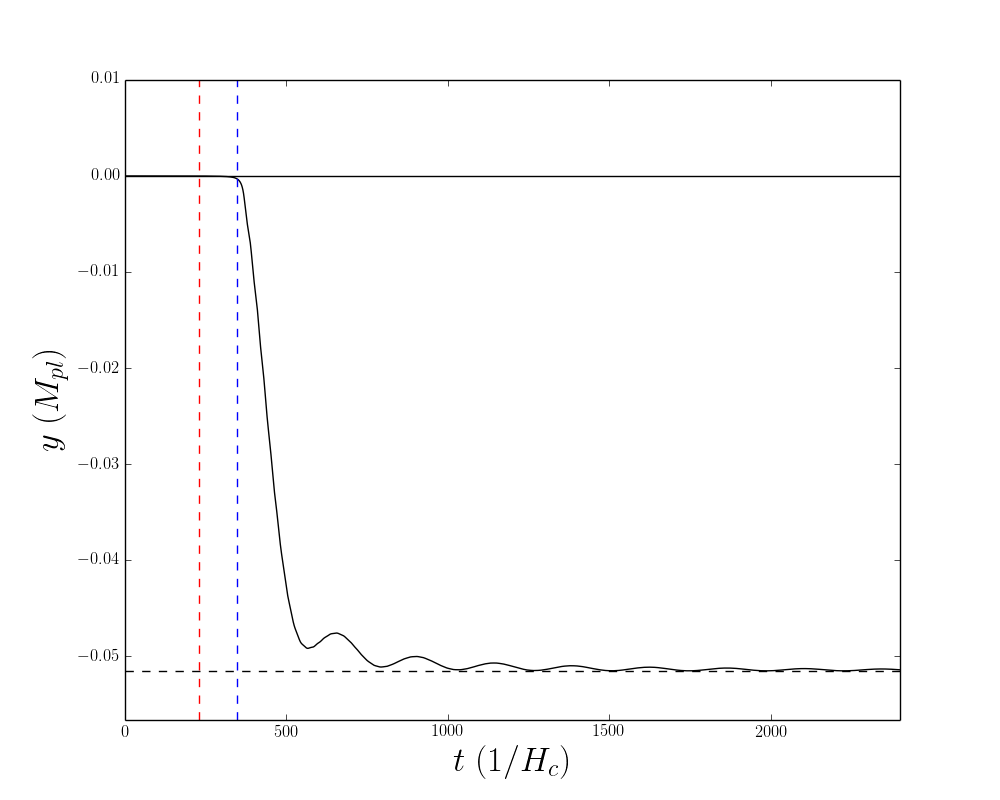}}
\end{subfigmatrix}
\caption{\footnotesize Evolution of the fields $\f$, $s$, and $y$ during and immediately after inflation. $H_c$ is the value of the Hubble parameter at the critical point. The values of $\f$ at the start of the last 60 $e$-folds, $\f_*$, and at the end of inflation, $\f_f$, are denoted by the vertical, dashed red and blue lines, respectively. The parameter values are $\k \simeq 4.5\times10^{-4}$, $\lam\simeq0.8$, and $v_{PS} \simeq 1.25\times10^{-2}\mpl\simeq 3\times 10^{16}\gev$. The initial conditions are discussed in~\cref{sec:reheating}.}
\label{fig:eoms}
\end{figure}

With this potential, we solve the coupled equations of motion
\begin{align}
\ddot{\f}+3H\dot{\f}+\prt_\f V &= 0, \label{eq:eoms}\\ 
\ddot{s}+3H\dot{s}+\prt_s V &= 0, \nn\\ 
\ddot{y}+3H\dot{y}+\prt_y V &= 0 \nn
\;.
\label{eq:eomss}
\end{align}
The behavior is shown in~\cref{fig:eoms}.
The initial conditions will be discussed in~\cref{sec:reheating} along with reheating. The $s$-dependent minimum value of $y$ is
\be
y_{\text{min}}(s) \simeq \frac{-\frac{\k m}{2\sqrt{2}}s^2}{m^2 + \frac{\l^2}{16\mpl^2}\lrp{s^2 - 2v^2_{PS}}^2 }
\;.
\label{eq:ymin}
\ee
Initially, $y$ is at its local minimum of zero and it remains near zero until the waterfall field is close to its global minimum. Once the waterfall is near its global minimum, the second term in the denominator of~\cref{eq:ymin} vanishes and we obtain the global minimum for $y$ discussed above, $-\k v^2_{PS}/(\sqrt{2}m)$.

Setting $y$ to zero and noting the relation $m/\mpl \ll \k \ll \l$, we obtain\footnote{Under the redefinitions, $\k\ra\l$, $\l\ra \sqrt{2}g$, and $v_{PS}\ra\sqrt{\xi}$,~\cref{eq:vinf} matches the potential in~\cite{Buchmuller:2014rfa} up to the $\half m^2\f^2$ term. In~\cite{Buchmuller:2014rfa} the flatness is lifted by quantum corrections rather than a small quadratic term.}
\be
V(\f,s,0) \simeq
\frac{\lam^2 v_{PS}^4 }{4} + \frac{1}{4}\lrp{\k^2 \f^2 - \lam^2 v_{PS}^2 }s^2 +
\frac{\lam^2}{16}s^4 + \half m^2\f^2
\label{eq:Vduring}
.\ee

\begin{figure}[t!]
\thisfloatpagestyle{empty}
\centering
\includegraphics[width=0.8\textwidth, trim=0ex 20ex 0ex 20ex, clip]{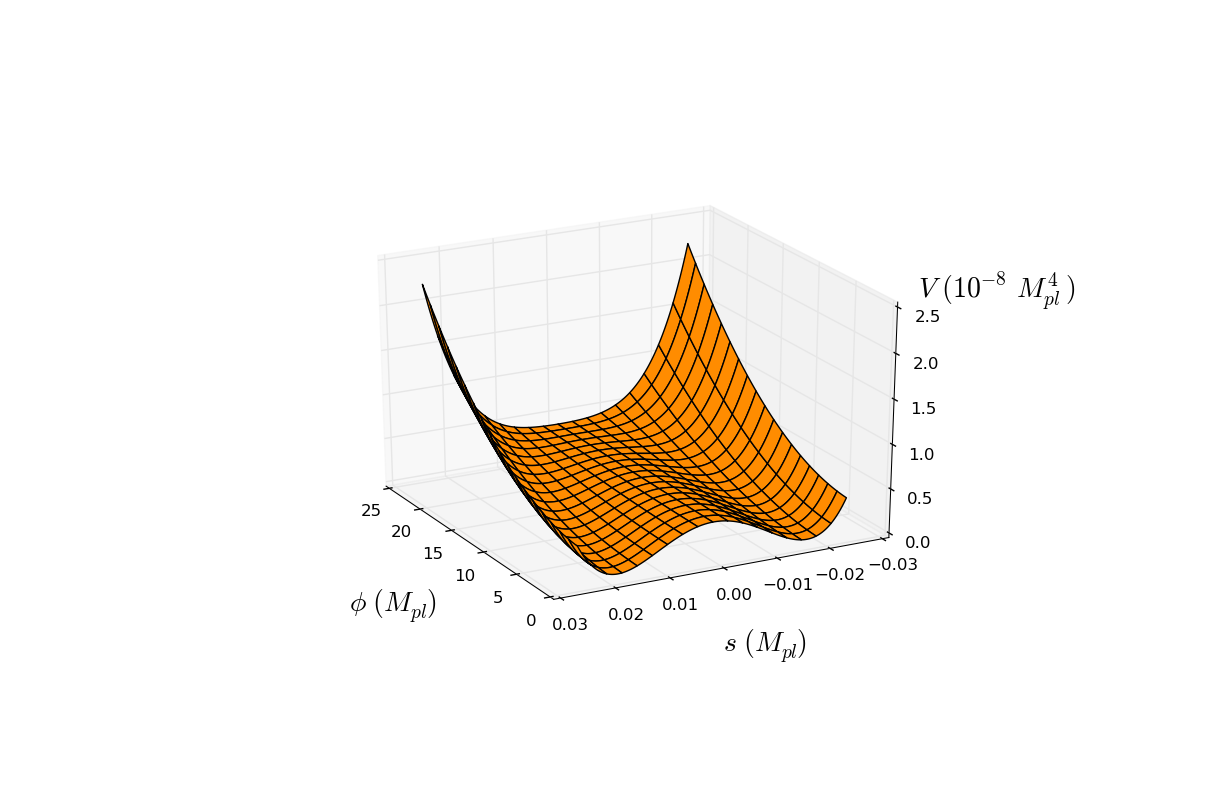}
\caption{\small The potential during inflation with y set to zero.}
\label{fig:3D}
\end{figure}

Initially, as the waterfall field $s$ is stabilized at zero, the potential is slightly quadratic in the inflaton direction which lifts the flatness above the critical point and allows the inflaton field $\f$ to approach its critical value, $\phi_c \equiv \frac{\lam v_{PS}}{\k}$. The shape of the potential is shown in~\cref{fig:3D}.

\begin{figure}[h!]
\thisfloatpagestyle{empty}
\centering
\includegraphics[width=0.5\textwidth, trim=0ex 0ex 0ex 0ex, clip]{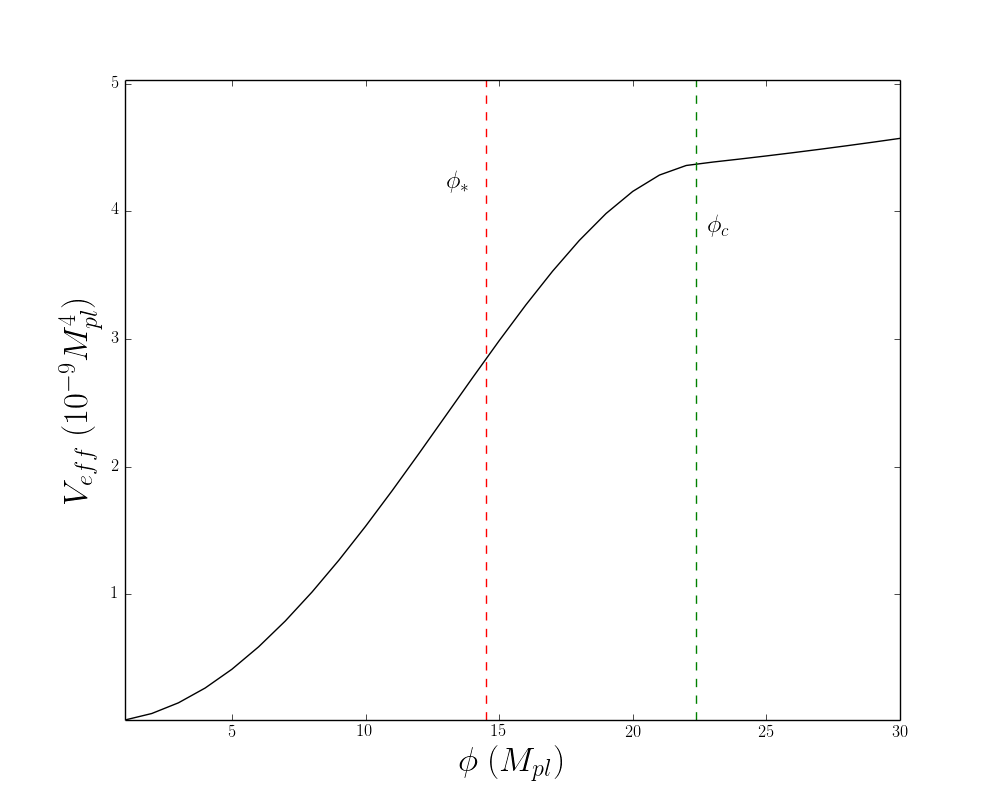}
\caption{\small The effective single-field potential during inflation. The critical point $\f_c$  is denoted by the vertical, dashed green line. The value of $\f$ at the start of the last 60 $e$-folds, $\f_*$, is denoted by the vertical, dashed red line. For values of $\f$ above $\f_c$, the potential is given by $V_0 = \frac{\l^2v^4_{PS}}{4}+\half m^2\f^2$.}
\label{fig:veff}
\end{figure}

Once the inflaton field attains subcritical values, the waterfall field quickly reaches its local $\f$-dependent minimum~\cite{Buchmuller:2014rfa},
\be
s^2_{\text{min}}(\phi)
= 2v_{PS}^2\lrp{1-\frac{\phi^2}{\phi_c^2}}
,\ee

and yields the effective potential that is relevant for the last 60 $e$-foldings of inflation,
\be
V_{eff}(\f) = \frac{\l^2 v_{PS}^4}{2}\frac{\phi^2}{\phi_c^2}\lrs{\lrp{1+\frac{m^2}{\k^2 v^2_{PS}}}-\frac{\phi^2}{2\phi_c^2}}
\simeq
\frac{\l^2 v_{PS}^4}{2}\frac{\phi^2}{\phi_c^2}\lrs{1-\frac{\phi^2}{2\phi_c^2}}
\;,
\label{eq:veff}
\ee
where $m^2/\k^2 v^2_{PS}\ll 1$. The effective potential is plotted in~\cref{fig:veff}.

The effective backreaction suppresses the steepness of the inflaton potential and causes the inflaton field to roll more slowly compared to chaotic inflation. It is easy to see from~\cref{eq:veff} that at $\f=\f_c$ the potential is approximately flat as in purely hybrid inflation models and at $\f \ll \f_c$ the potential is quadratic as in purely chaotic inflation. The interplay between the inflaton and waterfall fields in the subcritical inflation scenario therefore interpolates between these two regimes and can yield a tensor-to-scalar ratio consistent with the recent bound from BICEP2/\emph{Keck}~\cite{Array:2015xqh}.


\subsection{Cosmological Observables}

With $s=s_{\text{min}}(\phi)$, the slow roll parameters take their usual forms
\be
\e(\phi) = \half \lrp{\frac{V'_{eff}}{V_{eff}}}^2, \qquad
\eta(\phi) = \lrp{\frac{V''_{eff}}{V_{eff}}}
\;,\ee
where the prime denotes differentiation with respect to $\f$. The number of $e$-folds is then computed as
\be
N_e=\int^{\f_*}_{\f_f}\frac{1}{\sqrt{2\e(\f)}}
,\label{eq:phistar}
\ee
where $\f_*$ and $\f_f$ are the values of the inflaton field when the last 60 $e$-folds begins and when $\e=1$, respectively.~\cref{eq:phistar} can be solved to give $\f_*$ as a function of $N_e$,
\be
\f^2_* = \f^2_c\lrs{1-W_0\lrp{\Delta e^\Delta e^{-8N_e/\f^2_c}}}
,\ee
where $\Delta \equiv 1-\f^2_f/\f^2_*$ and $W_0$ is the principal branch of the Lambert $W$ function~\cite{Buchmuller:2014rfa}.

With the addition of the 95 GHz data from the \emph{Keck Array}, the BICEP2/\emph{Keck Array} experiments yield a tensor-to-scalar ratio $r < 0.09$ at 95\% confidence~\cite{Array:2015xqh}. The \emph{Planck} collaboration gives best fits for the scalar spectral index $n_s = 0.968\pm0.005$ and for the amplitude of the scalar power spectrum $A_s=(2.22\pm0.067)\times10^{-9}$~\cite{Ade:2015lrj}.

\begin{figure}[h!]
\thisfloatpagestyle{empty}
\centering
\includegraphics[width=0.6\textwidth, trim=0ex 10ex 10ex 15ex, clip]{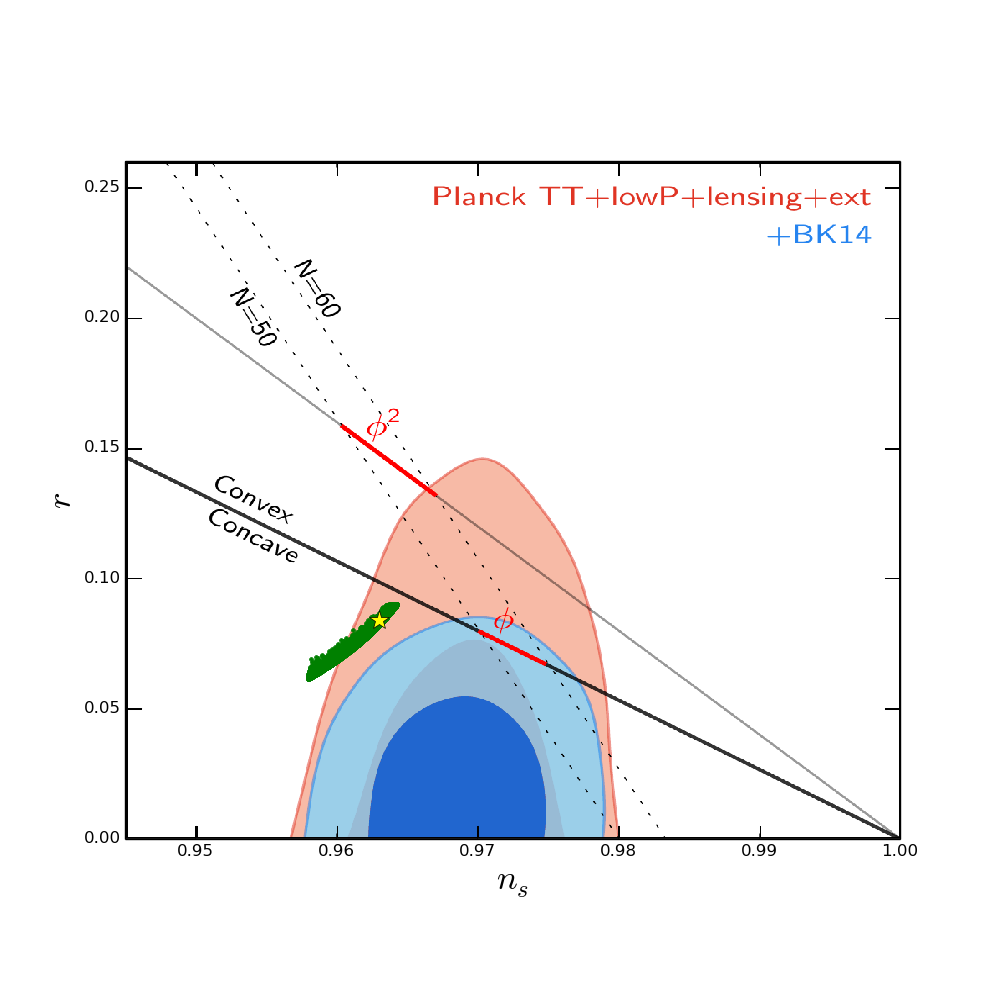}
\caption{\small The green points represent the result of our parameter scan and are overlayed on the best-fit plane found in~\cite{Array:2015xqh}. The yellow star represents our best fit point.}
\label{fig:overlay}
\end{figure}

Requiring $N_e=60$, we perform a scan over suitable values for the parameters $\k$, $\l$, $m$, and $v_{PS}$ and consider points that satisfy both the 2$\s$ upper bound on $r$ and the 2$\s$ bounds on $n_s$. In~\cref{fig:overlay} we overlay the result of the scan on the $r$-$n_s$ best-fit plane found in~\cite{Array:2015xqh}.

The best-fit point from the scan gives  $\k \simeq 4.5\times10^{-4}$, $\lam\simeq0.8$, $m=10^{-6}\mpl$ and $v_{PS} \simeq 1.25\times10^{-2}\mpl\simeq 3\times 10^{16}\gev$. With these parameter values, we find $\f_*=14.5~\mpl$ and the cosmological observables are computed to be
\be
r = 16 \e_* = 0.084,
\qquad n_s = 1 - 6\e_* +2\eta_* = 0.963
\qquad A_s = \frac{V_*}{24 \pi^2 \e_*} = 2.21\times10^{-9}
\;,\ee
where $\e_*=\e(\f_*)$ and $\eta_*=\eta(\f_*)$.

We note that with $\f_*/\f_c\simeq 2/3$ and $\l^2\simeq2/3$, ~\cref{eq:veff} gives $V_{eff}(\f_*)\simeq \lrp{\frac{2}{3} v_{PS}}^4$, and thus the energy scale during inflation is due directly to the GUT symmetry breaking scale, $v_{PS}$. The identification between the GUT scale and the energy scale during inflation was made in the context of $D$-term hybrid inflation with a Fayet-Iliopoulos term in~\cite{Buchmuller:2014dda}. 

\subsection{Mass spectrum}

To conclude this section, we discuss the resulting mass spectrum in the inflation sector after inflation ends. We express the scalar and fermion components of the superfields as
\be
\Phi\supset \lrp{\frac{a+i\f}{\sqrt{2}},\wt{\f}},\;
V^c\supset \lrp{\frac{\s+\sqrt{2}v_{PS}+i\tau}{\sqrt{2}},\wt{s}},\;
X\supset \lrp{\frac{x_r+i x_i}{\sqrt{2}},\wt{x}},\;
Y\supset \lrp{\frac{h -\k v_{PS}^2/\sqrt{2} m + i u}{\sqrt{2}},\wt{y}}\;.
\ee

Once the global minimum is reached, supersymmetry is restored and two massive chiral supermultiplets are formed. The mass eigenstates of the component fields are given as follows.   In the fermion sector, the states $\wt\f$ and $\wt y$ mix with $\wt x$ and $\wt s$ forming two Dirac mass eigenstates.   The mass matrix is given by  \be  \tilde m = ( \wt\f \;\; \wt x ) \left( \ba{cc} \kappa \ v_{ps} & m \\ \lambda \ v_{ps} & 0 \ea \right) \left( \ba{c} \wt s \\ \wt y \ea \right) . \ee  The mass eigenstates are found by diagonalizing the symmetric matrices
 \be \tilde m^T \ \tilde m =   \left( \ba{cc} (\kappa^2 + \lambda^2) \ v_{ps}^2 & \kappa \ v_{ps} \ m  \\ \kappa \ v_{ps} \ m  & m^2 \ea \right) \ee
 and  \be \tilde m \ \tilde m^T =   \left( \ba{cc} \kappa^2 \ v_{ps}^2 + m^2 & \kappa \ \lambda \ v_{ps}^2  \\  \kappa \ \lambda \ v_{ps}^2 & \lambda^2 \ v_{ps}^2 \ea \right). \ee

In the scalar sector we have the quadratic terms in the scalar potential for the real components of the fields $V^c$ and $Y$ given by
\be {\cal L} \supset \frac{1}{2} ( \sigma \;\; h) \left( \ba{cc} (\kappa^2 + \lambda^2) \ v_{ps}^2 & \kappa \ v_{ps} \ m  \\ \kappa \ v_{ps} \ m  & m^2 \ea \right) \left( \ba{c} \sigma \\ h \ea \right) .  \ee  And for the real components of the fields $\Phi$ and $X$ we have
\be {\cal L} \supset \frac{1}{2} ( a \;\; x_r) \left( \ba{cc} \kappa^2 \ v_{ps}^2 + m^2 & \kappa \ \lambda \ v_{ps}^2  \\  \kappa \ \lambda \ v_{ps}^2 & \lambda^2 \ v_{ps}^2 \ea \right) \left( \ba{c} a \\ x_r \ea \right) .  \ee  The mass squared matrices for the imaginary components of the fields are the same as for their real components.  We see that the spectrum is supersymmetric and the mass eigenvalues for the two different scalar sectors are identical with only the mixing angles being different.

The mass eigenvalues are given by
\be m_{1(2)}^2  =  \frac{1}{2} \ \left[(\k^2 + \lambda^2) \ v_{PS}^2 + m^2 +(-) \sqrt{ [(\k^2 + \lambda^2) \ v_{PS}^2 +  m^2]^2  - 4 \k^2 \ v_{PS}^2  m^2 } \right]. \ee  The mixing angles for $(\phi_1 \equiv \frac{\sigma + i \tau}{\sqrt{2}}, \phi_2 \equiv \frac{h + i u}{\sqrt{2}} )$  are given by
$\phi_i = O^T_{i j} \hat \phi_j$ (where $\hat \phi_j$ are the mass eigenstates), $O = \left( \ba{cc} \cos\theta & \sin\theta \\ - \sin\theta & \cos\theta \ea \right)$, and
\be \tan2\theta = \frac{2 \k v_{PS} m}{(\k^2 + \lambda^2) v_{PS}^2 - m^2} . \ee  The mixing angles for $(\chi_1 \equiv \frac{a + i \phi}{\sqrt{2}}, \chi_2 \equiv \frac{x_r + i x_i}{\sqrt{2}} )$  are given by
$\chi_i = \tilde O^T_{i j} \hat \chi_j$, $\tilde O = \left( \ba{cc} \cos\eta & \sin\eta \\ - \sin\eta & \cos\eta \ea \right)$, and
\be \tan2\eta = \frac{2 \k \l v^2_{PS} }{(\k^2 - \lambda^2) v_{PS}^2 + m^2} . \ee

Given the values of the parameters, $\lambda, \k$ and $v_{PS}, m$, we find the approximate form of the masses and mixing angles.  The mass eigenstates of the component fields are presented in~\cref{tab:scalars} and~\cref{tab:fermions} along with their masses and mixing angles. In the scalar sector, there is a very small amount of mixing between the scalar components of $\Phi$ and $X$ and also between the scalar components of $V^c$ and $Y$. We represent this slight misalignment between the interaction basis and the mass basis by simply placing a hat on the interaction-basis scalar fields (e.g. $\hat\f$ is the mass eigenstate that is approximately equal to $\f$.). In the fermion sector, the states $\wt\f$ and $\wt x$ mix with the same mixing angles as their scalar partners.  Likewise the states $\wt s$ and $\wt y$ mix.  The masses and mixing angles are given in~\cref{tab:scalars} and~\cref{tab:fermions}.

\begin{table}
\centering
\begin{tabular}{|c|c|c|c|}
\hline Interaction basis & Mass basis & Mass & Mixing angle   \\
\hline $\chi_1$, $\chi_2$ & $\hat \chi_1$, $\hat \chi_2$ & $m$ , $\lambda v_{PS}$ & $\t=\k m/(\l^2 v_{PS})$   \\
 $\phi_1$, $\phi_2$ & $\hat \phi_1$, $\hat \phi_2$ & $\l v_{PS}$ , $m$ &  $\eta =-\k/\l$ \\
\hline
\end{tabular}
\caption{\small Mass eigenstates, masses, and mixing angles of scalars. }
\label{tab:scalars}
\end{table}

\begin{table}
\centering
\begin{tabular}{|c|c|c|}
\hline Interaction basis & Mass basis & Mass   \\
\hline $\wt \f$, $\wt x$ & $\hat{\wt \f}$, $\hat{\wt x}$ & $m$ , $\l v_{PS}$  \\
 $\wt s$, $\wt y$ & $\hat{\wt s}$, $\hat{\wt y}$ &  $\l v_{PS}$ , $m$   \\
\hline
\end{tabular}
\caption{\small Mass eigenstates and masses of fermions.}
\label{tab:fermions}
\end{table}

\section{Matter Sector}

\subsection{Three family Pati-Salam model}

The matter sector of the theory is given by the superpotential  $\fw = \fw_I + \fw_{PS}$ with \bea \fw_{PS} = &  \fw_{neutrino} \  + \ \lambda Q_3 \ {\cal H} \ Q^c_3 \ + \ Q_a \ {\cal H} \ F_a^c \ + \ F_a \ {\cal H} \ Q_a^c & \\ & + \bar F_a^c \ ( M_0 \ \fo \ F_a^c \ + \ \phi_a \ {\cal O_{B-L}} \ Q_3^c \ + \ {\cal O_{B-L}} \ \frac{\theta_a \ \theta_b}{\hat M} \ Q_b^c \ - \ B_2 \ Q_a^c) & \nonumber \\ & + \bar F_a \ ( M_0 \ \fo \ F_a \ + \ \phi_a \ {\cal O_{B-L}} \ Q_3 \ + \ {\cal O_{B-L}} \ \frac{\theta_a \ \theta_b}{\hat M} \ Q_b \ + \ B_2 \ Q_a) & \nonumber ,\eea where  $ \{ Q_3, \ Q_a, \ F_a \} = (4, 2, 1,1), \;  \{ Q_3^c, \ Q^c_a, \ F_a^c \} = (\bar 4, 1, \bar 2, 1)$ with $a = 1,2$, a $D_4$ family index, ${\cal H} = (1, 2, \bar 2, 0)$, and the fields $\bar F_a,  \ \bar F_a^c$ are Pati-Salam conjugate fields.   The superpotential for the neutrino sector is given by
 \bea \fw_{neutrino}  &=& \Sb^c \ ( \lambda_2 \ N_a \ Q^c_a \ + \ \lambda_3 \ N_3 \ Q^c_3 )  \label{eq:neutrino}\nn \\ & & - \frac{1}{2} \ ( \l'_2 Y\ N_a \ N_a \ + \ \frac{\tilde \theta_a \ \tilde \theta_b}{\hat M} \ N_a \ N_b \ + \ \l'_3 Y \ N_3 \ N_3 )  \nonumber \\ &=&  \frac{\lambda_2^{2}}{2 \ M_1} (\Sb^c \ Q^c_1)^2 \ + \ \frac{\lambda_2^{2}}{2 \ M_2} (\Sb^c \ Q^c_2)^2 \ + \ \frac{\lambda_3^{2}}{2 \ M_3} (\Sb^c \ Q^c_3)^2   \label{eq:Wnu},\eea where  \be M_1 = \l'_2 Y, \ M_2 = \l'_2 Y + \frac{\tilde \theta_2^2}{\hat M}, \  M_3 = \l'_3 Y  \ee
and $\widetilde{\t}_1$ is taken to be zero.


  After expanding the waterfall field by its vev, the last line of Eq.~(\ref{eq:Wnu}) yields (with $\Sb^c \ra V^c/\sqrt{2}$ and $\l_1\equiv\l_2$)
  \be
  \frac{\l^2_i}{2 M_i}\lrp{\frac{\s+i\tau+\sqrt{2}v_{PS}}{2}}^2\nu^c_i \nu^c_i
  = \half M^i_R \nu^c_i \nu^c_i + \frac{h_i}{2}\lrp{\s+i\tau}\nu^c_i \nu^c_i
  \label{eq:sdecay}
  \ee
  plus terms quadratic in $\s,\tau$ with $M^i_R\equiv \frac{\l^2_i v^2_{PS}}{2 M_i}$ and $h_i \equiv \frac{\l^2_i v_{PS}}{\sqrt{2}M_i}$.

 Here we have identified $Y$ as one of the flavon fields. The ``right-handed" neutrino fields,  $N_a, \ N_3$ are PS singlets with charge $(1,1,1,1)$. The vev of $Y$ gives a heavy mass term for $N_a, \ N_3$ which are in turn integrated out to yield effective couplings between the waterfall field and the left-handed antineutrinos in $Q^c_a,\;Q^c_3$. Similar to the waterfall field, the scalar components of $Y$ also obtain a coupling to the left-handed antineutrinos
 \be
 \frac{h_i}{2}\lrp{\frac{m}{\k v_{PS}}}\lrp{h+iu}\nu^c_i \nu^c_i
 \;.
 \label{eq:hdecay}
 \ee

The fields $F_a, \ \bar F_a, \ F_a^c, \ \bar F_a^c$ are Froggatt-Nielson fields which are integrated out to obtain the effective Yukawa matrices.    We have defined the effective operators -  
\bea {\hat M^2 \ ({\cal O_{B-L}})^{\alpha i}}_{\beta j} &\equiv & - \frac{4}{3} \ {\delta^i}_j \ \bar {S^c}^{\gamma k} \ \lrp{{\delta^\alpha}_\gamma \ {\delta^\lambda}_\beta \ - \frac{1}{4} {\delta^\alpha}_\beta \ {\delta^\lambda}_\gamma}  \ S^c_{\lambda k}  \\ &=& {(B-L)^\alpha}_\beta \ {\delta^i}_j \ \frac{v_{PS}^2}{2}   \nonumber \eea
and  
\bea {\hat M^2 \ {\cal O}^{\alpha i}}_{\beta j} &\equiv&  \bar {S^c}^{\gamma k} \ \left[{\delta^\alpha}_\beta \ {\delta^i}_j \ {\delta^\lambda}_\gamma \ {\delta^l}_k + \tilde \alpha \ {\delta^\lambda}_\gamma \ \lrp{ {\delta^i}_k \ {\delta^l}_j - \frac{1}{2} {\delta^i}_j \ {\delta^l}_k}\right.  \\ & & \left. -\frac{4}{3} \ \tilde \beta \ {\delta^l}_k \ {\delta^i}_j \ \lrp{{\delta^\alpha}_\gamma \ {\delta^\lambda}_\beta  - \frac{1}{4} {\delta^\alpha}_\beta \ {\delta^\lambda}_\gamma}  \right] \ S^c_{\lambda l}  \nonumber \\ &=&  \lrs{{\mathbb{I}^{\alpha i}}_{\beta j} + \tilde \alpha \ {(T_{3R})^i}_j \ {\delta^\alpha}_\beta + \tilde \beta \ {(B-L)^\alpha}_\beta \ {\delta^i}_j } \ \frac{v_{PS}^2}{2}   \nonumber \\
&\equiv& \ \lrs{{\mathbb{I}^{\alpha i}}_{\beta j} + \alpha \ {(X)^{i \alpha}}_{j \beta} +  \beta \ {(Y)^{i \alpha}}_{j \beta} } \ \frac{v_{PS}^2}{2}  \nonumber   \eea

 where $X = 3 (B - L) - 4 T_{3R}$ commutes with $SU(5)$ and $Y = 2 T_{3R} + (B - L)$ is the SM hypercharge. The Froggatt-Nielson fields $F_a, \ \bar F_a, \  F_a^c, \ \bar F_a^c$ have a mass term given by $M_0 \ {{\fo}^{\alpha i}}_{\beta j}$.  The flavon fields $\{ \phi_a, \ \theta_a, \ \tilde \theta_a \}$ are doublets under $D_4$ and $B_2$ is a non-trivial $D_4$ singlet such that the product $B_2 * (x_1 y_2 - x_2 y_1)$ is $D_4$ invariant with $\{ x_a, \ y_a \}$ being $D_4$ doublets.  The $D_4$ invariant product between two doublets is given by $x_a y_a \equiv  x_1 y_1 + x_2 y_2$.  The flavon fields have zero charge under \zfr and are  assumed to get the non-zero VEVs -  $\{ \phi_{1,2}, \ \theta_2, \ \tilde \theta_2, \ \ B_2 \}$ and all others are zero.

Note, with the given particle spectrum and \zfr charges, we have the following anomaly coefficients, \be A_{SU(4)_C-SU(4)_C-\zfrmm} = A_{SU(2)_L-SU(2)_L-\zfrmm} = A_{SU(2)_R-SU(2)_R-\zfrmm} = 1 ({\rm mod}(2)) . \ee Thus the \zfr anomaly can, in principle, be canceled via the Green-Schwarz mechanism, as discussed in Ref. \cite{Lee:2010gv,Lee:2011dya}.  Dynamical breaking of the \zfr symmetry would then preserve an exact $R$-parity and generate a $\mu$ term, with $\mu \sim m_{3/2}$ and dimension 5 proton decay operators suppressed by  $m_{3/2}^2/M_{Pl}$.

\subsection{Yukawa matrices}

Upon integrating out the heavy Froggatt-Nielsen fields we obtain the effective superpotential for the low energy theory, \bea \fw_{LE} = & Y^u_{i j} \ q_i \ H_u \ u^c_j + Y^d_{i j} \ q_i \ H_d \ d^c_j + Y^e_{i j} \ \ell_i \ H_d \ e^c_j + Y^\nu_{i j} \ \ell_i \ H_u \ \nu^c_j  & \nonumber \\ &  + \frac{1}{2} \  M_{R_i} \nu^c_i \ \nu^c_i, \, {\rm for} \; i = 1,2,3 &  \nonumber \eea  where  \be M_{R_{1,2}} = \frac{\lambda_2^2 \ v_{PS}^2}{2 \ M_{1,2}}, \; M_{R_{3}} = \frac{\lambda_3^2 \ v_{PS}^2}{2 \ M_{3}} . \label{eq:masses} \ee  The Yukawa matrices for up-quarks, down-quarks, charged leptons and neutrinos given by (defined in Weyl notation with doublets on the left)\footnote{These Yukawa matrices are identical to those obtained
previously (see \cite{Dermisek:2005ij}) and analyzed most recently in \cite{Anandakrishnan:2012tj,Anandakrishnan:2014nea}.}
\bea
Y_u =&  \left(\begin{array}{ccc}  0  & \epsilon' \ \rho & - \epsilon \ \xi  \\
             - \epsilon' \ \rho &  \tilde \epsilon \ \rho & - \epsilon     \\
       \epsilon \ \xi   & \epsilon & 1 \end{array} \right) \; \lambda & \nonumber \\
Y_d =&  \left(\begin{array}{ccc}  0 & \epsilon'  & - \epsilon \ \xi \ \sigma \\
- \epsilon'   &  \tilde \epsilon  & - \epsilon \ \sigma \\
\epsilon \ \xi  & \epsilon & 1 \end{array} \right) \; \lambda & \label{eq:yukawaD31} \\
Y_e =&  \left(\begin{array}{ccc}  0  & - \epsilon'  & 3 \ \epsilon \ \xi \\
          \epsilon'  &  3 \ \tilde \epsilon  & 3 \ \epsilon  \\
 - 3 \ \epsilon \ \xi \ \sigma  & - 3 \ \epsilon \ \sigma & 1 \end{array} \right) \; \lambda &
 \nonumber \eea
with
\bea  \xi \;\; =  \;\; \phi_1/\phi_2, & \;\;
\tilde \epsilon  \;\; \propto   \;\;  (\theta_2/\hat M)^2,  & \label{eq:omegaD3} \\
\epsilon \;\; \propto  \;\; -\phi_2/\hat M, &  \;\;
\epsilon^\prime \;\; \sim  \;\;  ({ B_2}/M_0), \nonumber \\
  \sigma \;\; =   \;\; \frac{1+\alpha}{1-3\alpha}, &  \;\; \rho \;\; \sim   \;\;
  \beta \ll \alpha &  \nonumber \eea and
\bea Y_{\nu} =&  \left(\begin{array}{ccc}  0  & - \epsilon' \ \omega & {3 \over 2} \ \epsilon \ \xi \ \omega \\
      \epsilon'  \ \omega &  3 \ \tilde \epsilon \  \omega & {3 \over 2} \ \epsilon \ \omega \\
       - 3 \ \epsilon \ \xi \ \sigma   & - 3 \ \epsilon \ \sigma & 1 \end{array} \right) \; \lambda & \label{eq:yukawaD32}
  \eea
  with $\omega \;\; =  \;\; 2 \, \sigma/( 2 \, \sigma - 1)$ and a Dirac neutrino mass matrix given by
 \begin{equation} m_\nu \equiv Y_\nu \frac{v}{\sqrt{2}} \sin\beta.
 \label{eq:mnuD3}
  \end{equation}

From Eq.~(\ref{eq:yukawaD31}) and Eq.~(\ref{eq:yukawaD32}) one can see that the flavor hierarchies in the
Yukawa couplings are encoded in terms of the four complex parameters
$\rho, \sigma, \tilde \epsilon, \xi$ and the additional real ones
$\epsilon, \epsilon', \lambda$.
These matrices contain 7 real parameters and 4 arbitrary phases.
Note, the superpotential ($\fw_{PS}$) has many arbitrary parameters. However, the resulting effective Yukawa matrices have much fewer parameters and we therefore obtain a very predictive theory.   Also, the quark mass matrices accommodate the Georgi-Jarlskog mechanism, such that
$m_\mu/m_e \approx 9 \ m_s/m_d$.  This is a result of the $({\cal O_{B-L}})$ vev in the $B - L$ direction.


\section{Reheating \label{sec:reheating}}

After inflation ends, the inflaton and waterfall fields oscillate about their respective minima. Additionally, the $y$ field quickly reaches its minimum and also begins to oscillate.  In order to reach a radiation dominated era necessary for big bang nucleosynthesis (BBN), the energy stored in these fields must be transferred to decay products. Since the Hubble expansion causes radiation energy density to dilute at a faster rate than matter energy density does, we require that the fields be coupled to matter in order to avoid a matter dominated era during BBN. In this section we perform a simple calculation to determine the reheating temperatures from the inflaton and waterfall field decays.  Lastly we discuss the decay of $y$ oscillations. A full treatment of reheating and the evolution thereafter is left for future work.

In order to calculate the reheating temperatures it is necessary to determine the amount of energy density stored in the fields at the times when their respective decay rates become efficient. Because the fields are weakly coupled to one another and they are oscillating about their respective minima, we consider only the terms quadratic in the fields. We can therefore treat each field as having a separate energy density. After shifting the waterfall field by its global minimum, $s=\s+\sqrt{2}v_{PS}$, we find the separate $\f$ and $\s$ potential energy densities,
\be
V_\f \equiv \half \lrp{\k^2 v^2_{PS}+m^2} \f^2\;; \quad V_\s \equiv \half \l^2 v^2_{PS} \s^2
\;.\ee

Reheating occurs when the Hubble parameter becomes of order the decay rate. The Hubble parameter is a function of the total energy density,
\be
H^2 =\frac{\r_{tot}}{3 \mpl^2}
,\label{eq:hub}
\ee
where the total energy density is $\r_{tot}=\r_\f+\r_\s$ and the separate energy densities are
\be
\r_\f=\half\dot{\f}^2+V_\f \;;\quad
\r_\s=\half\dot{\s}^2+V_\s
.\ee

In our scenario, both the $\f$ and $\s$ fields contribute non-negligibly to the total energy density. As shown in~\cite{Buchmuller:2014rfa}, the waterfall field reaches its local inflaton-dependent minimum shortly after the waterfall transition, i.e. within a few e-folds. When solving the equations of motion to determine the values of the energy densities at the end of inflation, we therefore set the initial value of the waterfall field to be at its local minimum. It is left to determine the velocities of the fields near the critical point. For values of $\f$ above the critical point, the waterfall field is stabilized at zero and the tree level potential is given by $V_0 = \frac{\l^2v^4_{PS}}{4}+\half m^2\f^2$. The flatness of the potential is lifted slightly by the quadratic term and the velocity of the inflaton field immediately after reaching the critical point is determined from the slow-roll equation of motion of the inflaton,
\be
\dot{\f}_0
\equiv -\frac{1}{3H}\frac{\prt V_0}{\prt\f} \biggm\lvert_{\f_c}
=-\frac{2 m^2\mpl}{\sqrt{3} \k v_{PS}}
\;.\ee
For the parameter values given above, we find that $\dot{\f}_0/H_c\simeq-10^{-4}\f_c$ and the inflaton approaches its minimum very slowly.\footnote{In the usual hybrid inflation scenario, $\k\sim\fo(1)$ and $\dot{\f}_0/H_c $ is close to $\f_c$ causing inflation to end within a few $e$-folds.} Since the waterfall field quickly reaches its local minimum, as an approximation we take the initial velocity of the waterfall field to be
\be
\dot{s}_0\equiv \dot{\f}_0\frac{\prt s_{\text{min}}}{\prt \f}\biggm\lvert_{\f_c}
.\ee

With these initial conditions, we solved the equations of motion given by \cref{eq:eoms}. We find the energy densities at the end of inflation to be $\r^0_\f\approx 5\times10^{-11} \;\mpl^4$ and $\r^0_\s\approx 7\times10^{-14}\;\mpl^4$. Since $\r^0_\f$ dominates over $\r^0_\s$, we take the approximation, $H^2 \simeq\frac{\r_\f}{3 \mpl^2}$.

The inflaton field, $\f$, must convert its energy into matter. To achieve this we have introduced the operator:
\be
\fw_{\Phi}= \a \Phi \fh \fh .\ee
We find the decay rate into higgsinos (neglecting mixing) given by
\be
\Gamma_{\f\ra \wt{h}^0_u \wt{h}^0_d+\wt{h}^+_u \wt{h}^-_d} \approx \frac{\a^2}{8 \pi} \ m
\;.
\label{eq:phidecay}
\ee
The formula for the reheating temperature from $\f$ decays is given by the usual form
\be
T^\f_R = \lrp{\frac{90}{\pi^2 g_*}\Gamma_{\f}^2 \mpl^2}^{1/4}
\Ra
T^\f_R \simeq \lrp{2\times 10^{14}\gev}\a
\;,\ee where we have taken $g_* = 200$ and $\a\sim\fo(1)$.

Once the inflaton decays we are left with oscillations in the waterfall field, $\s$.\footnote{Note, we have implicitly assumed that
$\Gamma_{\f} \gg \Gamma_\s$ which is satisfied for $\alpha \sim 1$ and $h_1 \sim 10^{-6}$, as chosen below.}
The reheating temperature from $\s$ decays can now be calculated given the decay rate of $\s$ into matter. Note, the radiation energy density scales like the scale factor $a^{-4}$, while the energy in the $\s$ field scales like $a^{-3}$. Thus, as long as $\Gamma_{\f} \geq 10^6 \ \Gamma_\s$ the energy density of the universe when $\s$ decays is dominated by the energy in the $\s$ field.

 The rates for $\s$ to decay into two right-handed neutrinos or sneutrinos are
\be
\Gamma_{\s \ra \nu^c_i \nu^c_i}  =  \frac{h_i^2}{32 \pi} m_\s \lrp{1 - \frac{4 (M_R^i)^2}{m_\s^2}}^{3/2}
,\ee
\be
\Gamma_{\s \ra \widetilde{\nu}^c_i \widetilde{\nu}^c_i}  =  \frac{h_i^2}{32 \pi} m_\s \lrp{1 - \frac{4 (M_R^i)^2}{m_\s^2}}^{1/2}
.\ee
Since $M^i_R \ll m_{\s} \simeq 2.4 \times 10^{16}$ GeV, we can ignore the factors in parentheses and take the decay rate of $\s$ into matter to be
\be
\Gamma_{\s\ra \nu^c\nu^c+\wt\nu^c\wt\nu^c} =  \frac{h_1^2}{16 \pi} m_\s
,\label{eq:sigdecay}
\ee
where we have taken $h_1 \gg h_2,h_3$ for simplicity. To obtain an estimate of the reheating temperature from $\s$ decays, we take $M^1_R=7\times10^{10}\gev$ and $h_1=\sqrt{2}M^1_R/v_{PS}\simeq 3\times 10^{-6}$. Note that $\s$ can also decay into pairs of $\f$ or $a$, however, the rates for these decays are greatly suppressed compared to the decay rate into matter,
\be
\frac{\Gamma_{\s\ra\f\f+ a a}}{\Gamma_{\s\ra \nu^c\nu^c+\wt\nu^c\wt\nu^c}} \approx \frac{\k^4 v_{PS}^2}{h_1^2 \mpl^2}
\approx 10^{-6}
\;.\ee

We are now ready to calculate the reheating temperature.  The reheating temperature from $\s$ decays is given by
\be
T^\s_R = \lrp{\frac{90}{\pi^2 g_*} \Gamma_{\s}^2 \mpl^2}^{1/4}
\nn\ee
\be
\Ra
T^\s_R \simeq\lrp{3\times 10^{15}\gev}h_1
\;.\ee  
Using the value of $h_1$ from above we have \be \Gamma_\s \simeq 5 \times 10^3 \gev \ee and \be T^\s_R  \simeq 10^{10} \left(\frac{h_1}{3\times 10^{-6}}\right) \gev.\ee
Note, we want to choose a value for $h_1$ such that $T^\s_R < M^1_R$ so that the degree of CP violation produced when the right-handed neutrinos decay is not washed out from thermalization. Clearly this is easily done.  This is also consistent with the inflaton field decaying first and subsequently the waterfall field decaying with the final reheat temperature given by $T^\s_R$.

Finally, consider the $y$ field. The energy density after inflation stored in the physical field $h$ ($\rho^0_h \approx 10^{-17}\mpl^4$) is much less than the energy density stored in the waterfall field, $\s$. Furthermore, we see from~\cref{eq:sdecay} and~\cref{eq:hdecay}, that $\Gamma_{h\ra \nu^c\nu^c}=\lrp{m/\k v_{PS}}^2\Gamma_{\s\ra \nu^c\nu^c}\simeq 0.03\;\Gamma_{h\ra \nu^c\nu^c}$. Thus the energy density of the universe is still dominated by the energy in radiation due to $\s$ decay, when $h$ finally decays.

\section{Conclusions}
We have presented a Pati-Salam model of inflation and reheating which is consistent with recent cosmological data. The inflationary era is described by subcritical hybrid inflation which yields a tensor-to-scalar ratio consistent with the recent BICEP2/\emph{Keck} data. Furthermore, the energy scale during inflation is directly identified with the PS breaking scale.
Since the last 60 $e$-folds of inflation occur after the critical point, monopoles formed from the spontaneous breaking of the PS symmetry are diluted away. After inflation, the waterfall fields eventually decay into right-handed neutrinos at a reheat temperature below the mass of the lightest right-handed neutrino allowing for the possibilty of non-thermal leptogenesis via CP violation in the subsequent decays of the right-handed neutrinos. The model has also been extended to a three family model for fermion masses and mixing angles which reproduces results found previously in the literature~\cite{Dermisek:2005ij,Anandakrishnan:2012tj,Anandakrishnan:2014nea}.

In a future paper we intend to analyze reheating,  leptogenesis, and dark matter generation in the more complicated three family Pati-Salam model discussed earlier.  We know that since our right-handed neutrino masses are hierarchical with typical values of order $10^{10}, \ 10^{12}, \ 10^{14}$ GeV, a discussion of leptogenesis will require a detailed analysis of the production of a lepton asymmetry as well as washout effects.  Finally a discussion of dark matter will depend on the possible candidates.  With degenerate gaugino masses at the GUT scale, the lightest neutralino is bino-like.  This dark matter candidate typically over-closes the universe.  However with mirage mediation it has been shown that one can obtain a well-tempered dark matter candidate (see for example~\cite{Anandakrishnan:2013tqa}).  Any discussion of dark matter must also include the discussion of the SUSY breaking sector of the theory.  In particular, SUSY breaking must necessarily be decoupled from the inflationary sector such that the gravitino mass is less than the Hubble parameter during inflation, i.e. $m_{3/2} < H_{inf}$ (see Refs. \cite{Kallosh:2004yh,Kallosh:2011qk,Antusch:2011wu,Buchmuller:2015oma}).  In addition, the cosmological moduli \cite{Coughlan:1983ci,Ellis:1986zt,Banks:1993en,Banks:1995dt} and gravitino problems \cite{Weinberg:1982zq,Ellis:1984er,Ellis:1984eq} must be ameliorated.

\vspace{.2in}

{\bf Acknowledgements}

{\small S.R. and B.C.B. are partially supported by DOE grant DOE/DE-SC0011726.
We are also grateful to Archana Anandakrishnan and Zijie Poh for discussions.}

\bibliography{bibliography}

\providecommand{\href}[2]{#2}\begingroup\raggedright\begin{thebibliography}{10}

\bibitem{Kobayashi:2004ud}
T.~Kobayashi, S.~Raby, and R.-J. Zhang, ``{Constructing 5-D orbifold grand
  unified theories from heterotic strings},'' {\em Phys. Lett.} {\bf B593}
  (2004) 262--270,
\href{http://www.arXiv.org/abs/hep-ph/0403065}{{\tt hep-ph/0403065}}.

\bibitem{Kobayashi:2004ya}
T.~Kobayashi, S.~Raby, and R.-J. Zhang, ``{Searching for realistic 4d string
  models with a Pati-Salam symmetry: Orbifold grand unified theories from
  heterotic string compactification on a Z(6) orbifold},'' {\em Nucl. Phys.}
  {\bf B704} (2005) 3--55,
\href{http://www.arXiv.org/abs/hep-ph/0409098}{{\tt hep-ph/0409098}}.

\bibitem{Buchmuller:2014rfa}
W.~Buchmuller, V.~Domcke, and K.~Schmitz, ``{The Chaotic Regime of D-Term
  Inflation},'' {\em JCAP} {\bf 1411} (2014), no.~11, 006,
\href{http://www.arXiv.org/abs/1406.6300}{{\tt 1406.6300}}.

\bibitem{Buchmuller:2014dda}
W.~Buchmuller and K.~Ishiwata, ``{Grand Unification and Subcritical Hybrid
  Inflation},'' {\em Phys. Rev.} {\bf D91} (2015), no.~8, 081302,
\href{http://www.arXiv.org/abs/1412.3764}{{\tt 1412.3764}}.

\bibitem{Wieck:2014xxa}
C.~Wieck and M.~W. Winkler, ``{Inflation with Fayet-Iliopoulos Terms},'' {\em
  Phys. Rev.} {\bf D90} (2014), no.~10, 103507,
\href{http://www.arXiv.org/abs/1408.2826}{{\tt 1408.2826}}.

\bibitem{Kohri:2013gva}
K.~Kohri, C.~S. Lim, C.-M. Lin, and Y.~Mimura, ``{Hilltop Supernatural
  Inflation and SUSY Unified Models},'' {\em JCAP} {\bf 1401} (2014) 029,
\href{http://www.arXiv.org/abs/1309.4551}{{\tt 1309.4551}}.

\bibitem{Dvali:1994ms}
G.~R. Dvali, Q.~Shafi, and R.~K. Schaefer, ``{Large scale structure and
  supersymmetric inflation without fine tuning},'' {\em Phys. Rev. Lett.} {\bf
  73} (1994) 1886--1889,
\href{http://www.arXiv.org/abs/hep-ph/9406319}{{\tt hep-ph/9406319}}.

\bibitem{Jeannerot:2000sv}
R.~Jeannerot, S.~Khalil, G.~Lazarides, and Q.~Shafi, ``{Inflation and monopoles
  in supersymmetric SU(4)C x SU(2)(L) x SU(2)(R)},'' {\em JHEP} {\bf 10} (2000)
  012,
\href{http://www.arXiv.org/abs/hep-ph/0002151}{{\tt hep-ph/0002151}}.

\bibitem{Antusch:2010va}
S.~Antusch, M.~Bastero-Gil, J.~P. Baumann, K.~Dutta, S.~F. King, and P.~M.
  Kostka, ``{Gauge Non-Singlet Inflation in SUSY GUTs},'' {\em JHEP} {\bf 08}
  (2010) 100,
\href{http://www.arXiv.org/abs/1003.3233}{{\tt 1003.3233}}.

\bibitem{Buchmuller:2013dja}
W.~Buchmuller, V.~Domcke, K.~Kamada, and K.~Schmitz, ``{A Minimal
  Supersymmetric Model of Particle Physics and the Early Universe},''
\href{http://www.arXiv.org/abs/1309.7788}{{\tt 1309.7788}}.

\bibitem{Lazarides:1980cc}
G.~Lazarides, M.~Magg, and Q.~Shafi, ``{Phase Transitions and Magnetic
  Monopoles in SO(10)},'' {\em Phys. Lett.} {\bf B97} (1980)
87.

\bibitem{Carpenter:2014saa}
L.~M. Carpenter and S.~Raby, ``{Chaotic hybrid inflation with a gauged
  B–L},'' {\em Phys. Lett.} {\bf B738} (2014) 109--112,
\href{http://www.arXiv.org/abs/1405.6143}{{\tt 1405.6143}}.

\bibitem{Buchmuller:2012wn}
W.~Buchmuller, V.~Domcke, and K.~Schmitz, ``{Spontaneous B-L Breaking as the
  Origin of the Hot Early Universe},'' {\em Nucl. Phys.} {\bf B862} (2012)
  587--632,
\href{http://www.arXiv.org/abs/1202.6679}{{\tt 1202.6679}}.

\bibitem{Array:2015xqh}
{\bf BICEP2, Keck Array} Collaboration, P.~A.~R. Ade {\em et al.}, ``{BICEP2 /
  Keck Array VI: Improved Constraints On Cosmology and Foregrounds When Adding
  95 GHz Data From Keck Array},''
\href{http://www.arXiv.org/abs/1510.09217}{{\tt 1510.09217}}.

\bibitem{Ade:2015lrj}
{\bf Planck} Collaboration, P.~A.~R. Ade {\em et al.}, ``{Planck 2015 results.
  XX. Constraints on inflation},''
\href{http://www.arXiv.org/abs/1502.02114}{{\tt 1502.02114}}.

\bibitem{Lee:2010gv}
H.~M. Lee, S.~Raby, M.~Ratz, G.~G. Ross, R.~Schieren, K.~Schmidt-Hoberg, and
  P.~K.~S. Vaudrevange, ``{A unique $\mathbb{Z}_4^R$ symmetry for the MSSM},''
  {\em Phys. Lett.} {\bf B694} (2011) 491--495,
\href{http://www.arXiv.org/abs/1009.0905}{{\tt 1009.0905}}.

\bibitem{Lee:2011dya}
H.~M. Lee, S.~Raby, M.~Ratz, G.~G. Ross, R.~Schieren, K.~Schmidt-Hoberg, and
  P.~K.~S. Vaudrevange, ``{Discrete R symmetries for the MSSM and its singlet
  extensions},'' {\em Nucl. Phys.} {\bf B850} (2011) 1--30,
\href{http://www.arXiv.org/abs/1102.3595}{{\tt 1102.3595}}.

\bibitem{Dermisek:2005ij}
R.~Dermisek and S.~Raby, ``{Bi-large neutrino mixing and CP violation in an
  SO(10) SUSY GUT for fermion masses},'' {\em Phys. Lett.} {\bf B622} (2005)
  327--338,
\href{http://www.arXiv.org/abs/hep-ph/0507045}{{\tt hep-ph/0507045}}.

\bibitem{Anandakrishnan:2012tj}
A.~Anandakrishnan, S.~Raby, and A.~Wingerter, ``{Yukawa Unification Predictions
  for the LHC},'' {\em Phys. Rev.} {\bf D87} (2013), no.~5, 055005,
\href{http://www.arXiv.org/abs/1212.0542}{{\tt 1212.0542}}.

\bibitem{Anandakrishnan:2014nea}
A.~Anandakrishnan, B.~C. Bryant, and S.~Raby, ``{LHC Phenomenology of SO(10)
  Models with Yukawa Unification II},'' {\em Phys. Rev.} {\bf D90} (2014),
  no.~1, 015030,
\href{http://www.arXiv.org/abs/1404.5628}{{\tt 1404.5628}}.

\bibitem{Anandakrishnan:2013tqa}
A.~Anandakrishnan and K.~Sinha, ``{Viability of thermal well-tempered dark
  matter in SUSY GUTs},'' {\em Phys. Rev.} {\bf D89} (2014), no.~5, 055015,
\href{http://www.arXiv.org/abs/1310.7579}{{\tt 1310.7579}}.

\bibitem{Kallosh:2004yh}
R.~Kallosh and A.~D. Linde, ``{Landscape, the scale of SUSY breaking, and
  inflation},'' {\em JHEP} {\bf 12} (2004) 004,
\href{http://www.arXiv.org/abs/hep-th/0411011}{{\tt hep-th/0411011}}.

\bibitem{Kallosh:2011qk}
R.~Kallosh, A.~Linde, K.~A. Olive, and T.~Rube, ``{Chaotic inflation and
  supersymmetry breaking},'' {\em Phys. Rev.} {\bf D84} (2011) 083519,
\href{http://www.arXiv.org/abs/1106.6025}{{\tt 1106.6025}}.

\bibitem{Antusch:2011wu}
S.~Antusch, K.~Dutta, and S.~Halter, ``{Combining High-scale Inflation with
  Low-energy SUSY},'' {\em JHEP} {\bf 03} (2012) 105,
\href{http://www.arXiv.org/abs/1112.4488}{{\tt 1112.4488}}.

\bibitem{Buchmuller:2015oma}
W.~Buchmuller, E.~Dudas, L.~Heurtier, A.~Westphal, C.~Wieck, and M.~W. Winkler,
  ``{Challenges for Large-Field Inflation and Moduli Stabilization},'' {\em
  JHEP} {\bf 04} (2015) 058,
\href{http://www.arXiv.org/abs/1501.05812}{{\tt 1501.05812}}.

\bibitem{Coughlan:1983ci}
G.~D. Coughlan, W.~Fischler, E.~W. Kolb, S.~Raby, and G.~G. Ross,
  ``{Cosmological Problems for the Polonyi Potential},'' {\em Phys. Lett.} {\bf
  B131} (1983)
59.

\bibitem{Ellis:1986zt}
J.~R. Ellis, D.~V. Nanopoulos, and M.~Quiros, ``{On the Axion, Dilaton,
  Polonyi, Gravitino and Shadow Matter Problems in Supergravity and Superstring
  Models},'' {\em Phys.Lett.} {\bf B174} (1986)
176.

\bibitem{Banks:1993en}
T.~Banks, D.~B. Kaplan, and A.~E. Nelson, ``{Cosmological implications of
  dynamical supersymmetry breaking},'' {\em Phys.Rev.} {\bf D49} (1994)
  779--787,
\href{http://www.arXiv.org/abs/hep-ph/9308292}{{\tt hep-ph/9308292}}.

\bibitem{Banks:1995dt}
T.~Banks, M.~Berkooz, and P.~J. Steinhardt, ``{The Cosmological moduli problem,
  supersymmetry breaking, and stability in postinflationary cosmology},'' {\em
  Phys. Rev.} {\bf D52} (1995) 705--716,
\href{http://www.arXiv.org/abs/hep-th/9501053}{{\tt hep-th/9501053}}.

\bibitem{Weinberg:1982zq}
S.~Weinberg, ``{Cosmological Constraints on the Scale of Supersymmetry
  Breaking},'' {\em Phys. Rev. Lett.} {\bf 48} (1982)
1303.

\bibitem{Ellis:1984er}
J.~R. Ellis, D.~V. Nanopoulos, and S.~Sarkar, ``{The Cosmology of Decaying
  Gravitinos},'' {\em Nucl. Phys.} {\bf B259} (1985)
175.

\bibitem{Ellis:1984eq}
J.~R. Ellis, J.~E. Kim, and D.~V. Nanopoulos, ``{Cosmological Gravitino
  Regeneration and Decay},'' {\em Phys. Lett.} {\bf B145} (1984)
181.

\end{thebibliography}\endgroup

\bibliographystyle{utphys}

\end{document}